\documentclass[twocolumn,prl,aps,showpacs,superscriptaddress]{revtex4}
\usepackage{epsfig}
\usepackage{graphicx}
\graphicspath{{./Figures/}}

\begin{document}

\title{Quantum and Classical Dynamics of a BEC in a Large-Period Optical Lattice}

\author{J.~H.~Huckans} \email{huckans@phys.psu.edu} \affiliation{Joint Quantum Institute, National Institute of Standards and Technology, and University of Maryland, Gaithersburg, Maryland, 20899-8424, USA}
\author{I.~B.~Spielman} \affiliation{Joint Quantum Institute, National Institute of Standards and Technology, and University of Maryland, Gaithersburg, Maryland, 20899-8424, USA}
\author{B.~Laburthe~Tolra} \affiliation{\it Laboratoire de Physique des Lasers, Universit\'e de Paris 13, 93430 Villetaneuse, France}
\author{W.~D.~Phillips}
\author{J.~V.~Porto}

\affiliation{Joint Quantum Institute, National Institute of Standards and Technology, and University of Maryland, Gaithersburg, Maryland, 20899-8424, USA}

\date{\today}

\begin{abstract}
We experimentally investigate diffraction of a $^{87}$Rb Bose-Einstein condensate from a 1D optical lattice.  We use a range of lattice periods and timescales, including those beyond the Raman-Nath limit.  We compare the results to quantum mechanical and classical simulations, with quantitative and qualitative agreement, respectively.  The classical simulation predicts that the envelope of the time-evolving diffraction pattern is shaped by caustics: singularities in the phase space density of classical trajectories.  This behavior becomes increasingly clear as the lattice period grows.
\end{abstract}


\maketitle

\subsection{Introduction}

Modern atom optics~\cite{adams, burnett}, in particular the diffraction of atoms by standing waves of light~\cite{moskowitz,gould} provides a dramatic demonstration of the wave nature of atoms.  The advent of Bose-Einstein condensates (BEC), with their extremely narrow momentum distribution, has made resolving diffraction components straightforward in these systems~\cite{ovchinnikov}.  Most atom diffraction experiments with BECs focused on the regime where the diffracting standing wave has relatively few ``bound'' states~\cite{boundstate} or where the diffraction produces relatively few diffraction orders.  Under such conditions, the wave nature of the atoms is essential for describing the behavior of the system.  By contrast, when the optical potential has many bound states the quantum system can be well described by classical particle trajectories.  This latter regime has been theoretically investigated with such trajectories~\cite{berry1}, as well as from a matter wave perspective~\cite{bernhardt,janicke}.

In 1994, Janicke and Wilkens calculated the diffraction of cold atoms from a lattice potential and predicted a dramatic collapse and revival of the number of diffracted orders as the atoms coherently evolve in the lattice~\cite{janicke}.  Here we present the first experimental observation of this collapse and revival in both the quantum and classical regimes.  The envelope of the time-evolving diffraction pattern for a sinusoidal potential is dominated by caustics: singularities in the phase space density of classical trajectories~\cite{berry1}.  We investigate the quantum and classical regimes of atom wave diffraction by applying an optical standing wave (an optical lattice) to a BEC, and measuring the time evolution of the momentum distribution for a range of lattice periods.

While the phenomenon of atom diffraction is quantum mechanical, some early experiments used a classical trajectory approach to describe the observed channeling~\cite{salomon} and focussing~\cite{timp, mcclelland} of atoms by optical standing waves.  These experiments used thermal beams of atoms with a large momentum spread and measured the atomic position distribution within the standing wave.  Another early experiment used an optical standing wave to diffract laser-cooled atoms and observed the growth and subsequent collapse of the width of the diffraction pattern in the quantum regime~\cite{kunze}.  Later, the collapse and revival of a few diffraction orders was observed in the diffraction of a BEC from a lattice with only a few bound states~\cite{denschlag}.  Here we extend this earlier work by measuring the time-evolution of a BEC's momentum distribution in lattice potentials for a range of lattice periods (see Fig. 1).  We observe several collapses and revivals in both quantum and classical regimes.  We compare our results to the predictions of a single-particle quantum simulation and find excellent agreement.  We also develop a classical model which reproduces essential features of our data and provides physical insight into the evolution of the momentum distribution.

\subsection{Experimental Procedure}

We create a static 1D optical lattice at the intersection of two laser beams.  In the plane-wave approximation, the electric field of each beam  is $\vec{E}(\vec{r},t) = \hat e E_0 e^{i (\vec k \cdot \vec r - \omega t)} + c.c.$ where $\hat e$ is the polarization vector.  The combined electric field for both beams creates an optical potential for the atoms given by
\begin{equation}
\label{latticepotential}
U(z) = U_0\sin^2(\kappa_{\rm L}z).
\end{equation}
For a two-level system $U_0 = \hbar\Omega_0^2/\delta$,  and $\kappa_{\rm L} \equiv |\vec k_1 -\vec k _2| /2 = \pi/d$ is one-half of the magnitude of the reciprocal lattice vector, $M$ is the atomic mass, $d$ is the lattice period, $\Omega_0$ is the resonant, single-beam Rabi frequency, and $\delta<0$ is the detuning of the laser from atomic resonance (In this experiment, $U_0$ is nominally 30~$E_{\rm R}$ where $E_{\rm R} = \hbar^2 k^2/2M$ is the single-photon recoil energy.).  We vary $\kappa_{\rm L}$ by changing the angle between $\vec k_1$ and $\vec k_2$ and define $\hat{z}$ parallel to $\vec k_1 - \vec k_2$.  Near the energy minimum of each lattice site, the potential is nearly harmonic, $U_{\rm ho}(z) \approx M \omega_{\rm ho}^2 z^2 /2$, where $\omega^2_{\rm ho} = 2U_0\pi^2/Md^2$.  We choose the detuning $\left|\delta\right| \gg (\Omega_0,\Gamma$) (where $\Gamma$ is the natural linewidth of the atomic transition) so spontaneous emission is negligible for our experiment durations.
\begin{figure*} [p]
   \centering
   \includegraphics [width=6.5in] {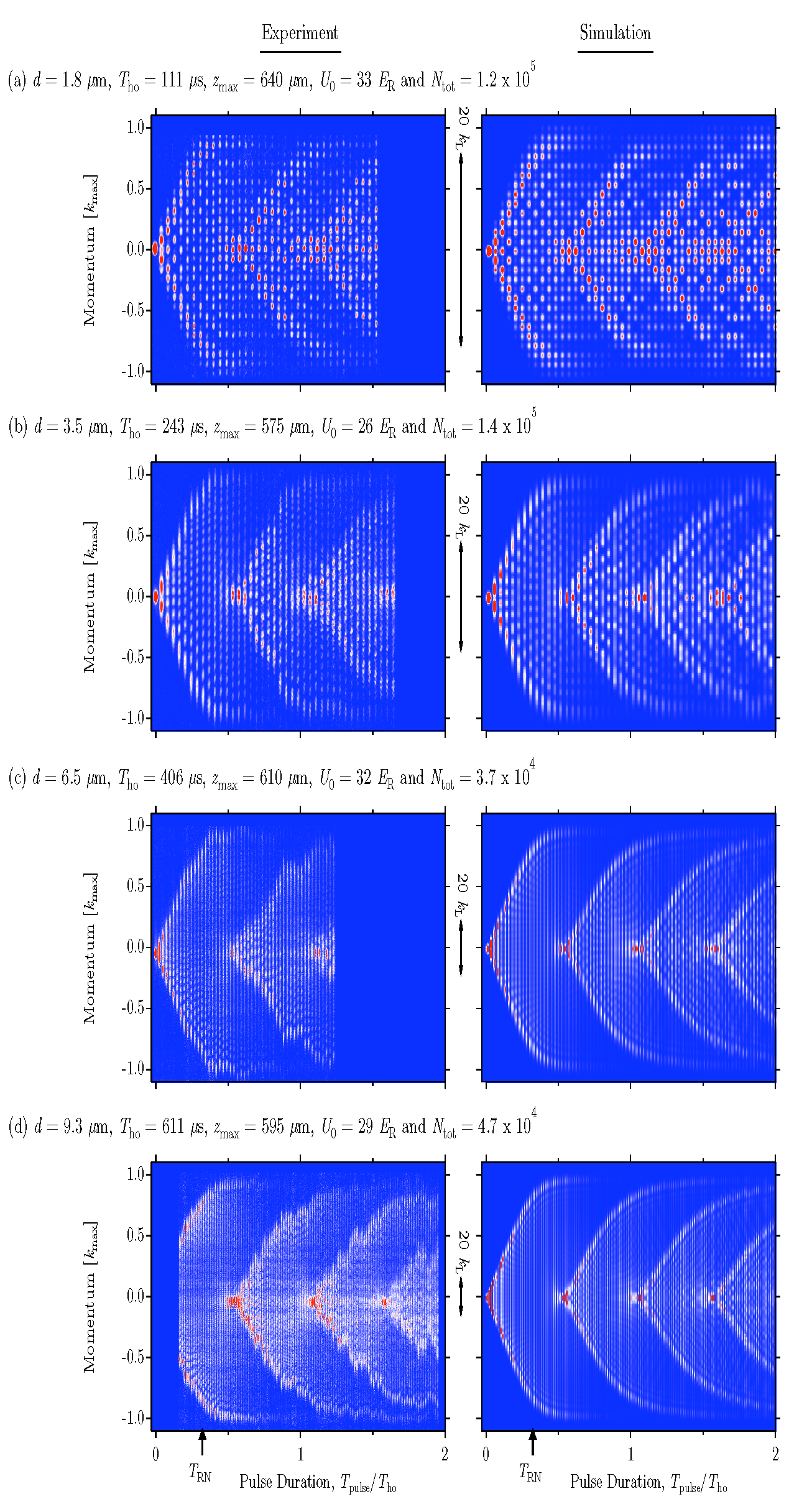}
   \caption{Concatenated absorption images of diffraction patterns, showing the evolving momentum distribution at four lattice periods $d$.  The momentum is scaled by $k_{\rm max} = z_{\rm max}M/t_{\rm TOF} \hbar$ where $z_{\rm max}$ is the measured maximum amplitude of the diffraction pattern after time-of-flight.  The calculated initial Thomas-Fermi radius in the lattice direction, $\hat{z}$, ranges from 26~$\mu$m to 29~$\mu$m, and in the transverse directions, from 6~$\mu$m to 10~$\mu$m.  The simulations were convolved with a Gaussian to model the effects of finite cloud size, mean-field-driven expansion and finite optical resolution.  $t_{\rm RN}$ is given by Eq. 2 and is the time at which the Raman-Nath approximation is expected to substantially fail.}
   \label{Figure:figex1}
\end{figure*}

We produce a nearly pure BEC with $N_{\mbox{\scriptsize tot}} = 5 \times 10^{4}$ to $14 \times 10^{4}$ atoms in the $(F,m_{F})=(1,-1)$ hyperfine state of $^{87}$Rb~\cite{peil03}.  We use a Ioffe-Pritchard trap with an oscillation frequency of $\nu_{\rm z} =8.2$~Hz in the weak direction and $\nu_{\rm x} = \nu_{\rm y} = 24$~Hz in the tight directions.  The optical lattice is suddenly applied to the magnetically-trapped atoms for a time $T_{\rm pulse}$.  The lattice periods $d$ used here, along with other relevant experimental parameters, are listed in Table I~\cite{stddev}.

\begin{table}[hbt]
\label{Experimental Data table}
\vspace{10pt}
\begin{tabular}{l|c|c|c|c}

                       &  (a)    & (b)        & (c)        & (d)        \\
\hline
$d$ ($\mu$m)           &  1.80(2) & 3.5(1) & 6.5(1) & 9.3(1) \\
$U_0$ ($E_{\rm R}$)    & 33(1)     & 26(2)    & 32(3)    & 29(3)    \\
$N_{\rm tot}$ ($10^4$) & 12(2)     & 14(5)    & 4(1)     & 5(3)     \\
$D_{\rm TF}$ ($\mu$m)  & 55(6)     & 57(6)    & 45(4)    & 46(5)    \\
$D_{\rm TF}/d$         & 31(3)     & 16(2)    & 7(1)     & 5.0(5)     \\
\end{tabular}
\caption{Experimental parameters for the four lattice periods investigated.  The Thomas-Fermi diameters $D_{\rm TF}$ along $\hat z$ were calculated using the measured atom numbers and known scattering length and trap frequencies.  Lattice depths were obtained by measuring the maximum kinetic energy of an atom during its evolution in a lattice.  The number of occupied lattice sites is approximately $D_{\rm TF}/d$.}
\end{table}

The lattice beams, linearly polarized perpendicular to the plane defined by the crossed beams, derive from a Ti:Sapphire laser operating at $\lambda=810$~nm (detuned below both $5S \rightarrow 5P$ transitions at 795~nm and 780~nm).  We constructed an ``accordion'' lattice allowing us to continuously vary the period of the diffracting potential~\cite{huckansthesis}.  Rotation of a galvanometer-controlled mirror causes the relative angle of the two beams to change while maintaining their intersection at the BEC.  The $e^{-2}$ radius of each beam is $\approx$~200~$\mu$m.  The lattice is turned on abruptly ($\lesssim 500\ {\rm ns}$) to its nominal depth of 30 $E_{\rm R}$, held constant for a variable time $T_{\rm pulse}$ and then turned off abruptly.  This constitutes a ``pulse'' of the lattice potential.  Immediately after the pulse, we release the atoms by turning off the magnetic trap in $\approx$~250~$\mu$s.  After the atom cloud expands for 20.2 ms, we record the spatial distribution of the atoms using resonant absorption imaging~\cite{cornell}.  Each image approximates the momentum distribution at the time of release.  Figure 1 shows a concatenated series of such images as a function of $T_{\rm pulse}$ at four different lattice periods.  Together, the  images reveal the evolving momentum distribution for each lattice period, which collapse and revive with the characteristic features predicted in Ref.~\cite{janicke}.

\subsection{Results and interpretation}

Figure 1a depicts the measured momentum distribution as a function of pulse duration for a BEC diffracted by a 1.8~$\mu$m lattice.  For small $T_{\rm pulse}$, the position of the apparent edge of the momentum distribution grows linearly with $T_{\rm pulse}$.  As expected, the distribution is composed of diffraction orders separated by $2\hbar \kappa_{\rm L}$.  At early times, our data are consistent with the diffraction predicted using the Raman-Nath approximation.  The Raman-Nath approximation can be viewed in a number of ways.  For example, it neglects the kinetic energy term in the single-particle Hamiltonian during application of the pulse.  This is equivalent to assuming that the only effect of the pulse is to impose a spatially periodic phase on the atomic wave function, with no effect on its amplitude profile.  This implies that the atoms move by a distance small compared to the lattice period $d$ during the pulse.  The Raman-Nath approximation is valid when
\
\begin{equation}
T_{\rm pulse} \ll t_{\rm RN} \equiv \frac {\hbar} {\sqrt {U_0  E_{\rm L}}} = \frac{T_{\rm ho}}{\pi},
\end{equation}
where $E_{\rm L} = \hbar^2 \kappa_{\rm L}^2/2M$ is the lattice recoil energy and $T_{\rm ho} = 2\pi/\omega_{\rm ho}$ is the harmonic oscillator period.  (Note that the lattice recoil in general differs from the photon recoil, but becomes equal to it when the lattice beams are counterpropagating.) In this approximation, the fractional atomic population of the $n^{\rm th}$ diffraction order is $J_n^2(U_0 T_{\rm pulse}/2\hbar)$ where the $J_n$ are Bessel functions of the first kind.

\begin{figure}[htbp]
   \centering
   \includegraphics [width=8.5cm]{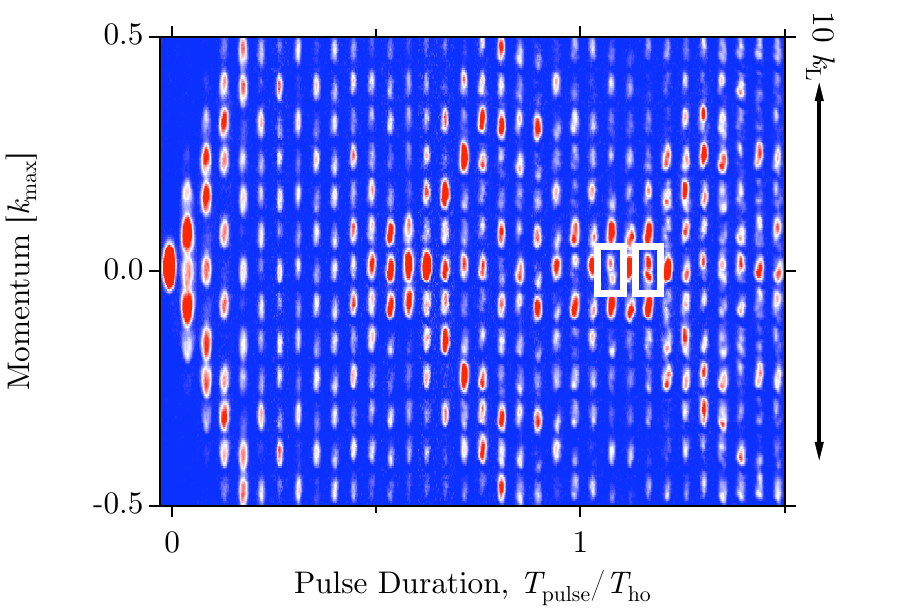}
   \caption{Splitting of diffraction orders (1.8 $\mu$m lattice period) for pulse durations beyond the Raman-Nath regime.  The effect is more pronounced the longer the pulse duration and two examples are indicated with white boxes.}
   \label{Figure:figex2}
\end{figure}

Figure 1a shows that as the pulse duration increases beyond $t_{\rm RN}$, the apparent edge of the momentum distribution is bounded by a maximum momentum $\hbar k_{\rm max}$.  We use this observed value of $\hbar k_{\rm max}$ to determine the lattice depth $U_0 = \hbar^2 k_{\rm max}^2/2M$.  The numerical calculations verify the accuracy of this identification to within 3~$E_{\rm R}$.  (The validity of the identification relies on the lattice being deep enough to support many bound states and the rise and fall times of the pulse being short compared to $T_{\rm ho}$.)  Once the edge of the distribution reaches $k_{\rm max}$, it gradually fades and a new outgoing edge appears near $k=0$, shortly after $T_{\rm ho}/2$.  Diffraction orders re-emerge as the new edge moves outward to higher momentum and the process approximately repeats.  At each collapse, a large fraction of the population returns to the lowest orders.  Collapses repeat at times which are approximate multiples of $T_{\rm ho}/2$ .

The collapses and revivals can be understood from a classical model where atoms initially at rest in the sinusoidal potential oscillate; those starting near the bottom of the potential oscillate at approximately $\omega_{\rm ho}$.  As we will see below, the suddenness of each collapse results from the anharmonicity of the potential and can also be understood in our classical model.  Because of the anharmonicity, no collapse is total, and there remains a sizable occupation of the higher momentum orders at the collapse point.  This collapse and revival always occurs at times greater than $t_{\rm RN}$ and illustrates a complete breakdown of the Raman-Nath approximation.

Figures 1b-d depict the momentum evolution for longer lattice periods.  While each evolution is similar, there are differences.  For $d > 1.8 \mu$m, the diffraction orders overlap; indeed, mean-field-driven expansion of the individual orders can lead to such overlap for all times-of-flight.  The apparent discreteness in the momentum distribution in Fig 1b, does not represent individual momentum orders but rather a modulated momentum-envelope over several unresolved orders.  As the lattice period $d$ increases and the spacing between the momentum orders decreases, the momentum distribution appears as a continuos curve.  Also, the point of collapse occurs increasingly close to $T_{\rm ho}/2$ as $d$ increases.

The quantum simulations shown in the right-hand column of Fig. 1 reproduce these features of our experiment.  They depict the {\it in-situ} momentum distribution after $T_{\rm pulse}$.  We numerically solve the time-dependent Gross-Pitaevskii (GP) equation during $T_{\rm pulse}$ including the mean-field interactions~\cite{nointeractions}.  We assume that the solution factors into a time-independent radial wavefunction and a time-dependent axial wavefunction.  We treat the axial component using a 1D GP equation with effective interaction strength $g_{\rm 1D} = 4 g_{\rm 3D} / 3\pi R_{\rm x} R_{\rm y}$ where $g_{\rm 3D} = 4\pi\hbar^2 a_s/M$ is the 3D interaction strength~\cite{spielman}.  Here $R_{\rm x}$ and $R_{\rm y}$ are the Thomas-Fermi radii in the directions perpendicular to the lattice, and $a_s$ is the s-wave scattering length for our $^{87}$Rb atoms.  Figure 1 shows the good agreement between these simulations and our experimental data. The simulation stops at the end of each pulse and we plot the momentum distribution (before the time-of-flight).

A detail of the experimental results, absent in the simulations, is the visible splitting of some diffraction orders for pulse durations beyond the Raman-Nath regime (see Fig. 2).  The effect becomes more pronounced the longer the pulse duration.  The effect implies spatial structure  larger than the lattice spacing and it becomes more pronounced the longer the pulse duration. One possible explanation is that the approximate periodic translational symmetry of the lattice is gradually compromised as the inhomogeneous mean-field interaction increases over time due to the atoms accelerating radially toward the cloud center because of the dipole force arising from the laser beam profile.  If this were the mechanism, it would explain why we do not see the order splittings in the simulation since radial dynamics are neglected.

Certain salient features reproduced by the numerical simulations can also be understood on the basis of simple arguments, both quantum and classical.  In the following two sections, we first give a single-particle quantum mechanical argument which explains the collapse and revival periods and their deviations from $T_{\rm ho}/2$.  We continue with a classical explanation for the sudden collapse of the higher momentum orders and their subsequent revival.

\subsection{Quantum Mechanical Analysis}

To understand the atomic evolution after sudden application of an optical lattice, it is useful to decompose the initial BEC state into the relevant eigenstates of the lattice potential with energies $E_n$.  We assume an initial state with momentum eigenvalue zero, equivalent to an infinite-extent BEC, projected onto the Bloch states (with zero quasimomentum) of 1D sinusoidal lattices having various periods.  Only even parity Bloch states are occupied since the initial wavefunction is symmetric.  We calculate the projection onto all significantly occupied states and find that the bound states contain the vast majority of the population.  Figure 3 shows the populations of all Bloch states up to the first unbound state, which include more than 99.5\% of the population.

\begin{figure}[htb!]
   \centering
   \includegraphics [width=8.5cm]{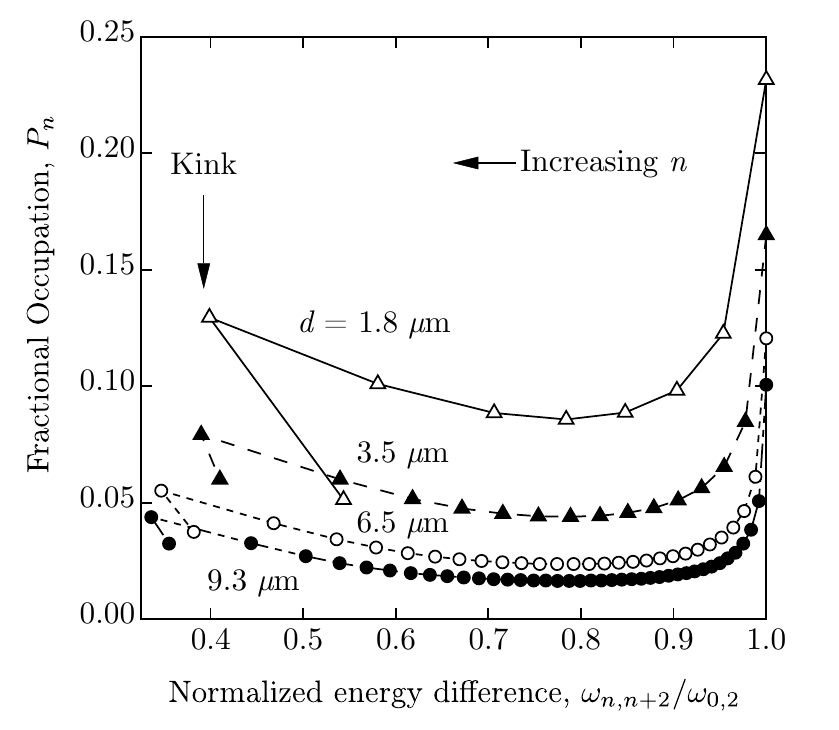}
   \caption{Calculated projections of a uniform wavefunction onto even-parity, zero-quasimomentum Bloch states of a 30~$E_{\rm R}$ lattice with periods of 1.8 $\mu$m, 3.5 $\mu$m, 6.5 $\mu$m, and 9.3 $\mu$m.  Each point corresponds to one such Bloch state.  The horizontal axis gives the energy separation to that state's even-parity, higher energy neighbor, normalized to the separation of the lowest energy pair, and the vertical axis gives the fractional occupation of the state.  The difference between consecutive energy levels first decreases with increasing $n$, then increases as states become unbound.  This explains the sharp kinks at the last data point (the first unbound state).}
   \label{Figure:figex2}
\end{figure}

\begin{figure}[htbp]
   \centering
   \includegraphics[width=8.5cm]{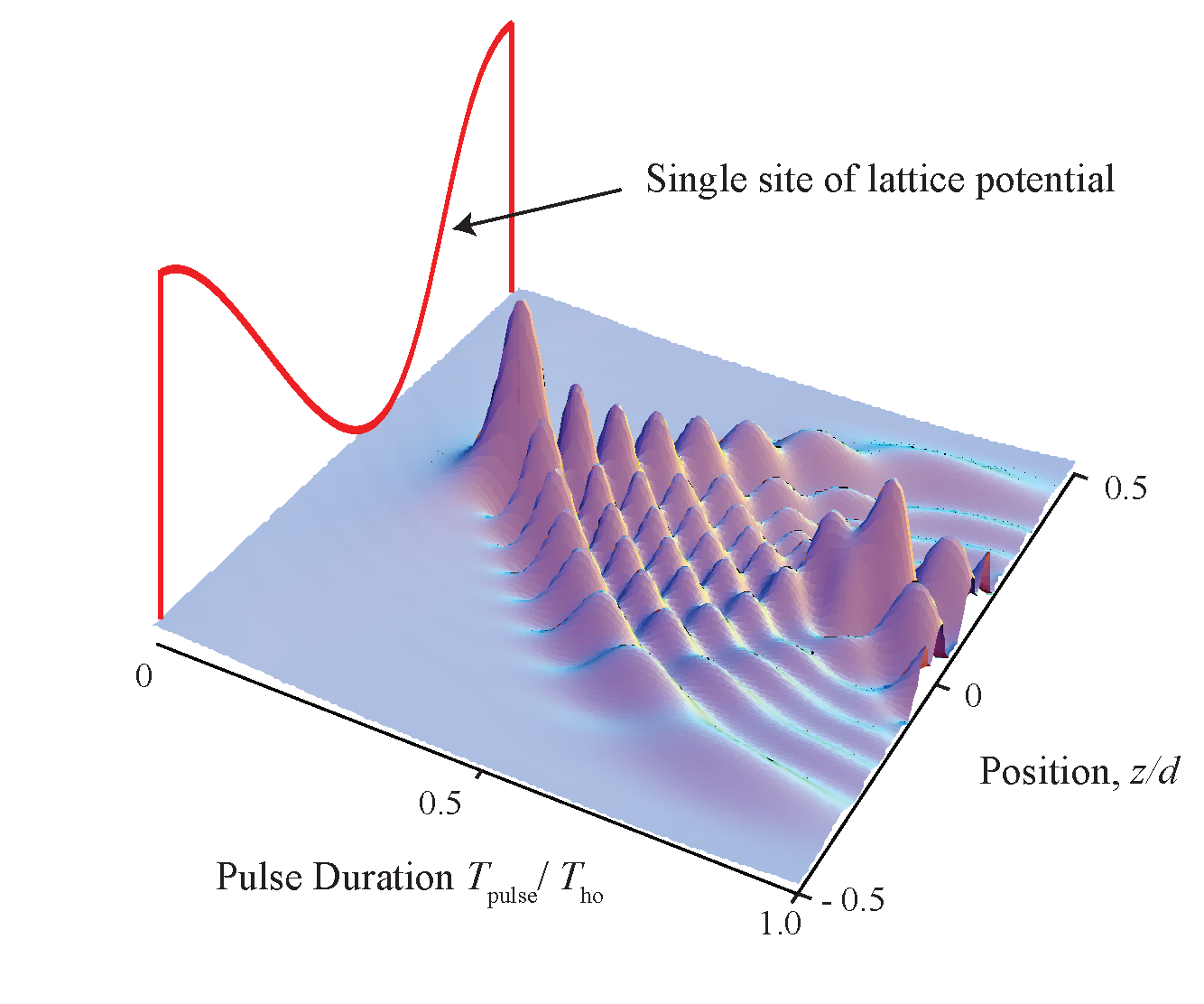}
   \caption{Calculated evolution of the probability density $|\Psi(z,t)|^2$ within a single site of a lattice (potential shown in red) with period $d = 1.8~\mu$m and depth 30~$E_{\rm R}$.  Positions $z/d= -0.5$ and $0.5$ correspond to the edges of the well.  When $T_{\rm pulse}/T_{\rm ho} \approx 0.25$ there is a large increase in density at the center of the well, at which time a large number of high momentum states are occupied.}
   \label{Figure:figex3}
\end{figure}

The number of significantly occupied eigenstates grows with increasing lattice period.  The increase of occupied bound states corresponds to the system becoming increasingly classical.  In the deep-lattice limit, the number of bound states in a sinusoidal potential scales as $(U_0/E_{\rm L})^{1/2}$.  Since the lattice recoil energy $E_{\rm L} \propto d^{-2}$, the number of bound states in a lattice grows linearly with the lattice period at fixed total depth.  For example, a lattice with a depth of 30 $E_{\rm R}$ formed by two counter-propagating beams has a period $d= \lambda/2$, a depth of 30~$E_{\rm L}$, and 4 bound states.  However, a 30 $E_{\rm R}$ lattice formed by two beams intersecting at 87 mrad $\approx 5^\circ$ has a period $d = 9.3~\mu$m, a depth of $15.8 \times 10^3~E_{\rm L}$, and $\approx 80$ bound states.

The phase of the $n^{\rm th}$ Bloch state evolves independently as $\omega_n t$ where $\omega_n = E_n/\hbar$.  In a harmonic potential, the frequency differences would be multiples of the harmonic frequency and all even eigenstates would rephase with period $T_{\rm ho}/2$.  The anharmonic potential of a single lattice site leads to a non-uniform spacing of the energy levels, so the system never perfectly rephases.  Nevertheless, the wavefunction approximately rephases such that a large fraction of the population is in diffraction orders close to zero momentum at times close to $T_{\rm ho}/2$ . (See Fig.~4 where a wide spread in position corresponds to the collapse of the momentum distribution as seen in Fig.~1.)

\subsection{Classical Analysis}

\begin{figure}[htbp]
   \centering
   \includegraphics[width=9cm]{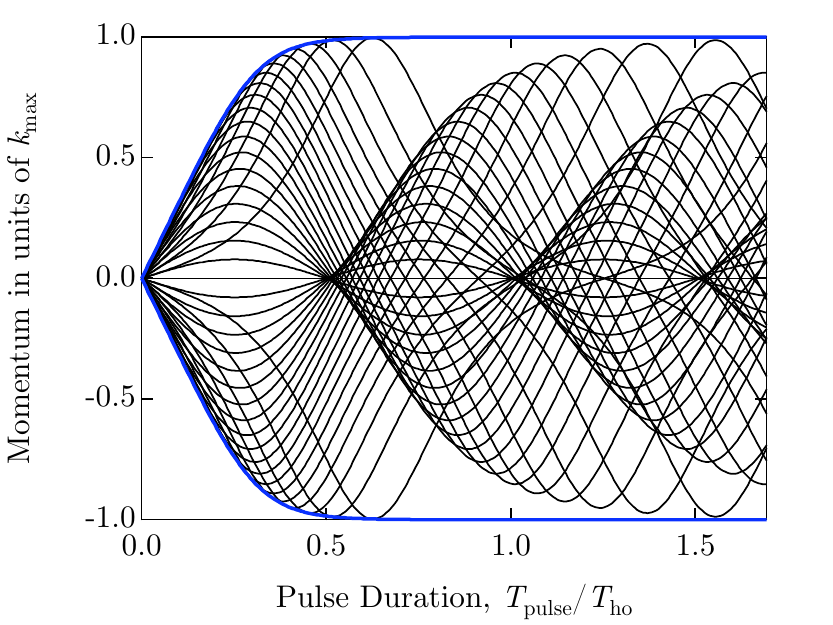}
   \caption{Classical motion of particles in a sinusoidal potential.  The curves
correspond to the trajectories of particles starting at different
points in the potential, each with zero initial momentum, $p_z=0$.
The smallest amplitude oscillations correspond to the smallest
period, $T_{\rm ho}$.  The
oscillation period diverges for particles starting increasingly near
the top of the sinusoidal potential.  The blue curve indicates the first caustic, a classical analog to the first apparent high density feature in our quantum system.}
   \label{Figure:figex4}
\end{figure}

Many aspects of this quantum mechanical system can be understood classically, in some cases quantitatively.  We study the corresponding classical system by calculating the trajectories of an ensemble of particles initially at rest and distributed uniformly in a sinusoidal potential (see Fig. \ref{Figure:figex4}).  Each trajectory reaches a turning point where the velocity returns to zero.  For small amplitude oscillations in the sinusoidal potential, the motion is nearly harmonic and atoms reach these turning points approximately concurrently.  Larger amplitude oscillations are increasingly anharmonic, leading to increasing times to reach the turning points.  Figure \ref{Figure:figex4} displays several momentum-trajectories of initially motionless classical particles evolving in a sinusoidal potential~\cite{sinediffraction}.

\begin{figure}[htbp]
   \centering
   \includegraphics[width=9cm]{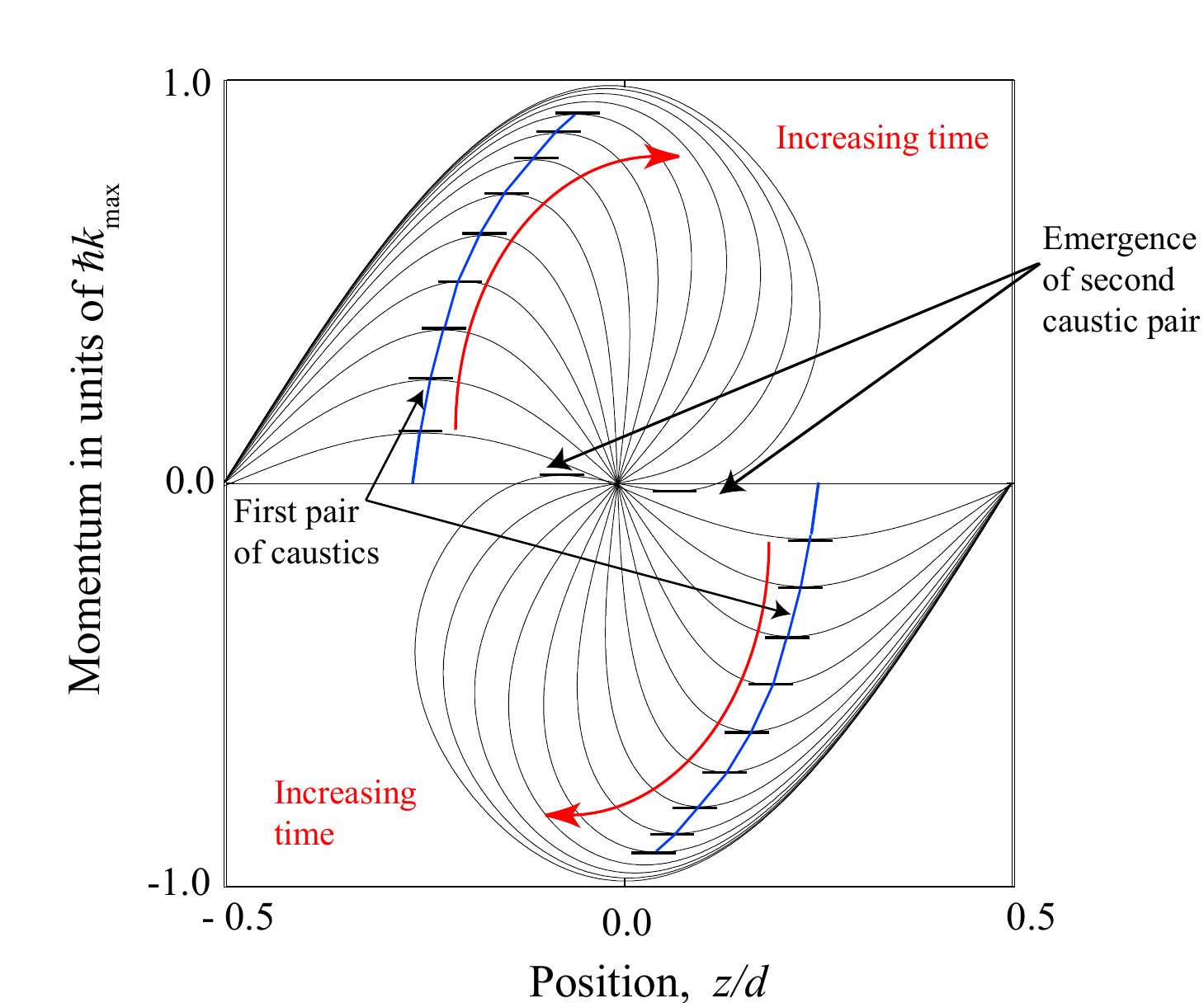}
   \caption{Classical phase space evolution of atoms in a single well of a sinusoidal potential.  Each black curve depicts the phase space distribution at a specific time.  The lips of a single well are located at positions $z=\pm d/2$.  The phase space evolution is characterized by fixed points at the lips and rotation approaching the harmonic frequency near $z=0$.}
   \label{Figure:figex5}
\end{figure}

Insight can be gained from this classical picture.  For
example, the suddenness of the collapse can be understood~\cite{berry1} by referring to a single-well phase space portrait
plotting position vs. momentum as shown in Fig.~\ref{Figure:figex5}:
Starting with a uniform distribution at rest ($p_z=0$), the
distribution rotates clockwise about the origin.  At the origin, the
period of rotation is $2\pi/\omega_{\rm ho}$.  Points farther from
the origin rotate more slowly about the origin.  The rotational
period diverges for atoms nearest the top of the sinusoidal
potential, $z=\pm d/2$.  The classical analog of our measured
momentum distribution is the projection of this evolving phase space
distribution onto the momentum axis.  The horizontal tangents of the
phase space distribution (indicated in the figure) project to
singularities in the momentum distribution.  The locus of these
tangent points is referred to as a caustic.  As the distribution
approaches the first turning point, the two horizontal tangents are
near the maximum momentum; there are no singularities near $p_z= 0$.
When the central part of the distribution crosses the turning point
(at $T_{\rm ho}/2$), another pair of caustics suddenly emerges from
the origin.  This explains the asymmetry in the momentum evolution;
the caustics only appear after $T_{\rm ho}/2$.  (The time evolution
is symmetric only for a perfectly harmonic potential.)

In a quantum or wave-optics system, diffraction softens the divergence of the underlying classical caustics.  The sequence in Fig. 1a-d shows, for the first time in a matter wave system, a progression from diffractive caustics~\cite{thom, arnold1, arnold2,trinkaus} toward classical caustics: the diffractive structure is manifest for the shortest period lattice and nearly invisible in the longest.

\subsection{Conclusion}

We measured the collapse and revival of the diffraction pattern of a BEC exposed to a pulsed 1D optical lattice.  Our results are found to be in good agreement with the predictions of the time-dependent Gross-Pitaevskii equation.  In addition, we employed a classical model that captures essential features and adds physical insight to the evolving diffraction pattern.  For long lattice periods, bound states proliferate and we observed classical behavior in the long-time evolution of the momentum distribution.  We captured, for the first time, the emergence of both diffractive and classical caustics in ultra-cold atom systems interacting with optical lattices.

We gratefully acknowledge helpful discussions with Paul Julienne, Ennio Arimondo, Mikkel Andersen, and Vincent Boyer.  This work was partially supported by ARDA, NASA, NRL and IBS acknowledges the support of the NRC Postdoctoral Fellowship Program.


\begin{thebibliography}{99}
\bibitem{boundstate} We define a bound state for this periodic potential to be a zero quasimomentum Bloch state whose energy is within the depth of the potential.
\bibitem{stddev} All uncertainties reported here are 1 standard deviation combined statistical and systematic uncertainties.
\bibitem{nointeractions} In the absence of interactions, the results are nearly indistinguishable from those shown in Fig. 1.
\end{thebibliography}
\end{document}